# Coherent and incoherent magnons induced by strong ultrafast demagnetization in thin permalloy films

Anulekha De,[1, *] Akira Lentfert,[1] Laura Scheuer,[1] Benjamin Stadtmüller,[1, 2]
Georg von Freymann,[1, 3] Martin Aeschlimann,[1] and Philipp Pirro[1, †]

[1]*Department of Physics and Research Center OPTIMAS,
Rheinland-Pfälzische Technische Universität Kaiserslautern-Landau, 67663 Kaiserslautern, Germany*
[2]*Institute of Physics, Johannes Gutenberg University Mainz, 55128 Mainz, Germany*
[3]*Fraunhofer Institute for Industrial Mathematics, 67663 Kaiserslautern, Germany*

(Dated: August 14, 2023)

Understanding spin dynamics on femto- and picosecond timescales offers new opportunities for faster and more efficient spintronic devices. Here, we experimentally investigate the coherent spin dynamics after ultrashort laser excitation by time-resolved magneto optical Kerr effect (TR-MOKE) in thin $Ni_{80}Fe_{20}$ films. We provide a detailed study of the magnetic field and pump fluence dependence of the coherent precessional dynamics. We show that the coherent precession lifetime increases with the applied external magnetic field which cannot be understood by viscous Gilbert damping of the coherent magnons. Instead, it can be explained by nonlinear magnon interactions and by the change in the fraction of incoherent magnons. This interpretation is in agreement with the observed trends of the coherent magnon amplitude and lifetime as a function of the exciting laser fluence. Our results provide a new insight into the magnetization relaxation processes in ferromagnetic thin films, which is of great importance for further spintronic applications.

## I. INTRODUCTION

The microscopic mechanism of laser-induced magnetization dynamics in femto-, pico- and nanosecond timescales remains still a challenge in condensed matter physics. Using the time-resolved magneto-optical Kerr effect (TR-MOKE), one can directly address the processes responsible for the excitation and relaxation of a magnetic system on their characteristic timescales [1–4]. The pioneering work of Beaurepaire et al. in 1996 on femtosecond laser-induced ultrafast demagnetization opened up a new avenue for ultrafast manipulation of the magnetization in magnetic materials [5]. Only a few years later, experiments could show that the ultrafast laser pulses also generate coherent magnons, manifested as precessional dynamics on the nanosecond time scale. The precessional region in magnetic materials is a key parameter for encoding and transferring information in spintronic devices. This regime allows the study of the magnetic anisotropy, damping, and precession frequency of different dynamic modes in continuous thin films and patterned nanostructures [2–4, 6–10]. However, the transition between the two regimes of the magnetization dynamics, i.e., the ultrafast demagnetization and the coherent precessional motion on the other side, is very crucial and raises intriguing questions. Many efforts have been made to understand the leading mechanisms as well as to explore the characteristic time scales. For example, van Kampen, et al. [2] demonstrated using TR-MOKE that an optical pump pulse can induce coherent uniform spin precession in a ferromagnet. The excitation of these magnons was explained by a transient change in magnetic anisotropy. Later, several works were concerned with exploring the precessional relaxation mechanisms in ferromagnets after ultrafast excitation [11–15]. A recent work showed the excitation of the precession dynamics in ferromagnets with two non-collinear optical pulses, that can affect precessional relaxation mechanisms and damping [16].

In this article, we take a closer look at the relaxation of the coherent precession induced by ultrafast demagnetization and study in detail the decay time of the measured coherent oscillations as a function of magnetic field and pump fluence. We use femtosecond amplified laser pulses to excite and detect magnetization dynamics in thin permalloy films, including ultrafast demagnetization, fast and slow remagnetization and precessional dynamics. We observe an unusual magnetic field dependence of the precessional relaxation time which is not in accordance with the expectations from the Gilbert model. Extrinsic contributions such as two-magnon scattering, magnetic anisotropy and spin pumping [11, 17–22] could also affect the magnetization relaxation process and the damping, but we show that these effects can be neglected in out case. Instead, we can relate our observations to nonlinear magnon interactions and the variable contribution of the incoherent magnon background after excitation by a femtosecond optical pulse. Our findings are of general importance for the interpretation of the coherent dynamics measured in similar experiments after ultrafast stimuli.

## II. MATERIALS AND METHODS

Permalloy ($Ni_{80}Fe_{20}$, Py) films of 5 nm and 2.8 nm thickness were deposited on MgO substrates by molecu-lar beam epitaxy (MBE) technique in an ultrahigh vacuum chamber. The samples were capped with a 3 nm thick layer of $Al_2O_3$ to protect them from environmental degradation, oxidation and laser ablation during the pump-probe experiment using femtosecond laser pulses.

---

* [ade@rptu.de](mailto:ade@rptu.de)
† [ppirro@rptu.de](mailto:ppirro@rptu.de)

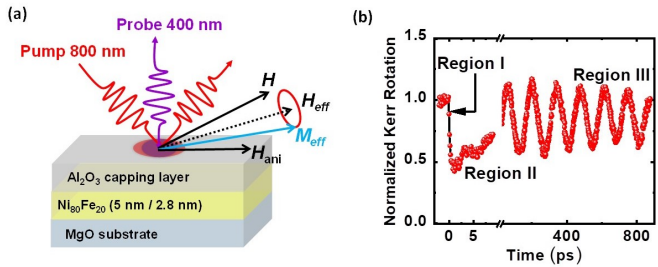

FIG. 1. (a) Schematic diagram of the measurement geometry. (b) Typical experimental TR-MOKE data showing different temporal regimes of the magnetization dynamics for 5 nm Py sample measured at $\mu_0 H = 113$ mT and $F = 4.6$ mJ cm$^{-2}$.

The magnetization dynamics were measured using a TR-MOKE setup based on a two-color, non-collinear optical pump-probe technique. A schematic diagram of the experimental geometry is shown in Fig. 1(a). In this experiment, we use the fundamental output of an amplified femtosecond laser system with wavelength $\lambda = 800$ nm, repetition rate of 1 kHz, and pulse width of $\sim 35$ fs (Libra, Coherent Inc.) as the pump pulse, while its second harmonic with $\lambda = 400$ nm is used to probe the dynamics. The probe is normally incident on the sample, while the pump is incident obliquely ($\sim 30°$) with respect to the surface normal. During the measurements, we have applied a magnetic field inclined at a small angle of $\sim 15°$ to the sample plane. The inclination of the magnetization provides a finite out-of-plane (OOP) demagnetization field which is transiently modified by the pump pulse, inducing a coherent precession of the magnetization [1, 2]. This symmetry breaking is important to obtain the same starting phase of the precessional motion for each laser pump pulse to avoid that the loss of the precession signal in a TR-MOKE experiment averaged over thousands of pump-probe cycles. For a completely in-plane configuration of the external magnetic field, we observe a complete reduction of the precessional signal [see Supplemental Materials]. To obtain the intrinsic magnetic response, we performed the measurements for two opposite magnetization directions of the sample and extracted the pure magnetic response from the difference of the two resulting Kerr signals. This is done to eliminate any nonmagnetic signal, i.e., any signal that does not depend of the direction of the sample magnetization [23]. We have used a specially designed photodetector connected to a lock-in amplifier to measure the dynamic Kerr rotation signal. All measurements have been performed under ambient condition and room temperature.

## III. RESULTS AND DISCUSSION

Several processes occur when a femtosecond laser pulse interacts with a ferromagnetic thin film in its saturation condition. First of all, the magnetization of the system is partially or completely lost within hundreds of femtoseconds, which is known as ultrafast demagnetization [5]. This is generally followed by a fast recovery of the magnetization within sub-picoseconds to a few picoseconds and a slower recovery within hundreds of picoseconds, known as the fast and slow remagnetization. The slower recovery is accompanied by a precession of the magnetization. On a much longer time scale of a few nanoseconds, the magnetization returns to its initial equilibrium, which can be described phenomenologically by the Gilbert damping [24]. Figure 1(b) shows the representative Kerr rotation data of the 5 nm Py sample for pump fluence $F = 4.6$ mJ cm$^{-2}$ and $\mu_0 H = 113$ mT consisting of three temporal regions of the magnetization dynamics, i.e., the ultrafast demagnetization, the fast remagnetization followed by the slow remagnetization superposed with the damped precession within the time window of 900 ps. The slow remagnetization is mainly due to heat diffusion from the lattice to the substrate and the surroundings. Region I is characterized by the demagnetization time $\tau_M$, region II is characterized by the fast remagnetization time $\tau_E$. For region III, we characterize the dynamics by the precession frequency $f$ and the precessional relaxation time $\tau_d$.

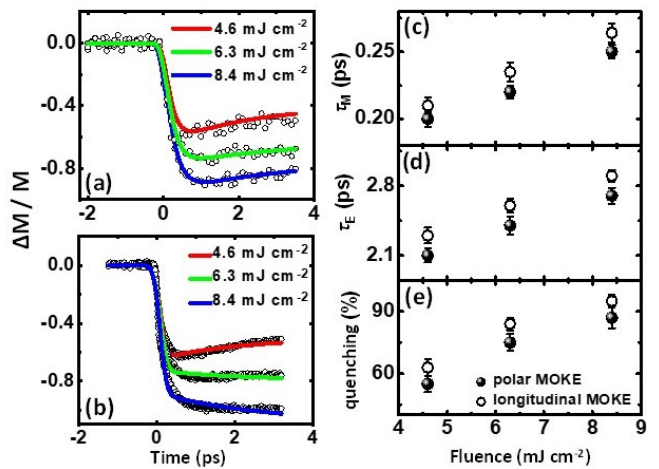

FIG. 2. Ultrafast demagnetization traces at different pump fluences for 5 nm Py sample in (a) polar (b) longitudinal MOKE geometry. (c) Demagnetization times ($\tau_M$) vs. pump fluence (d) Fast remagnetization times ($\tau_E$) vs. pump fluence and (e) quenching vs. pump fluence measured in both polar and longitudinal geometry. Solid circles and open circles represent the data corresponding to polar and longitudinal MOKE respectively.

We start our discussion with the ultrafast demagnetization and fast remagnetization processes of our system to study their dependence on the pump laser fluence and the external magnetic field. In general, the ultrafast demagnetization studies are performed in the longitudinal MOKE geometry. However, to measure the coherent precession of the magnetization, we need to perform measurements in a polar MOKE geometry with a slightly out-of-plane inclined external magnetic field [1, 2]. Therefore, we first check whether there is a significant difference in the laser-induced ultrafast demagnetization in the two MOKE geometries. Figure 2(a) and (b) show the ultrafast demagnetization traces obtained for 5 nm Py in the polar and longitudinal MOKE geometries, respectively. The pump fluence is varied between 4.6 - 8.4 mJ cm$^{-2}$ by varying the power of the pump pulse. We restrict ourselves to the low pump fluence regime to avoid sample damage, and the probe fluence is kept constant at a very low value ($\sim$ few $\mu$W) to avoid any additional contribution to the spin dynamics by probe excitation. The ultrafast demagnetization traces for both geometries show qualitatively similar trends. However, minor discrepancies between the quantitative values obtained from two different geometries arise due to slight differences in the pump spot sizes. The amplitude of the maximum quenching of the Kerr rotation signal increases almost linearly with the laser fluence. Closer inspection of the traces also reveals an increase in the $\tau_M$ and $\tau_M$ with increasing fluence. To quantify this increase, we fit our demagnetization traces with a phenomenological thermodynamic model, the so-called three-temperature model (3TM) [25], which is obtained by solving the energy rate equation between three different degrees of freedom, e.g. electron, spin and lattice, under low pump fluence conditions [see Supplemental Materials].

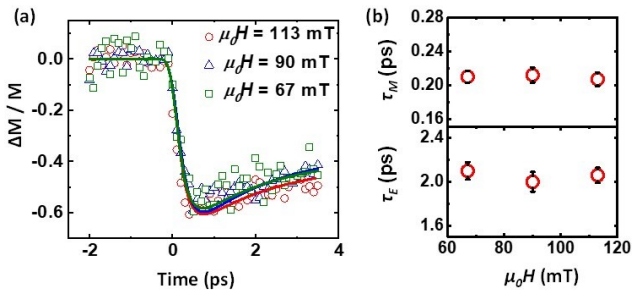

FIG. 3. (a) Ultrafast demagnetization traces for 5 nm Py sample measured at different values of magnetic field and for fixed $F = 4.6$ mJ cm$^{-2}$. (b) upper panel: Demagnetization times ($\tau_M$) vs. magnetic field and lower panel: Fast remagnetization times ($\tau_E$) vs. magnetic field.

The fluence-dependent behavior of $\tau_M$, $\tau_E$ and quenching in both MOKE geometries, as shown in Fig. 2(c), (d) and (e), respectively, is an indication of the spin-flip process-dominated ultrafast demagnetization in our systems [26–28]. The values of $\tau_M$ extracted from our experiments are on the same time scale as previous reports [29], and are too large to represent superdiffusive transport driven demagnetization [30]. These values are slightly larger in the longitudinal geometry compared to the polar geometry. However, they do not vary significantly with the applied external magnetic field (as shown in Fig. 3), indicating that, as expected, the comparatively small variations in Zeeman energy as well as the small change in magnetization direction associated with the change in magnetic field strength do not affect the ultrafast magnetization dynamics.

After quantifying the ultrafast demagnetization and fast remagnetization dynamics (region I and II), we turn to the region III, which is characterized by the coherent precessional magnetization dynamics induced by the pump laser pulse. These dynamics in the GHz- range are generally described by the phenomenological Landau-Lifshitz-Gilbert (LLG) equation [24],

$$\frac{\mathrm{d}\vec{M}}{\mathrm{d}t} = -\gamma \vec{M} \times \left[\mu_0 \vec{H}_{\mathrm{eff}} - \frac{\alpha}{\gamma M_{\mathrm{S}}} \frac{\mathrm{d}\vec{M}}{\mathrm{d}t}\right] \quad (1)$$

where $\gamma$ is the gyromagnetic ratio, $M_{\mathrm{S}}$ is the saturation magnetization, $\alpha$ is the Gilbert damping constant, and $\vec{H}_{\mathrm{eff}}$ is the effective magnetic field consisting of several field components. The first term on the right side of Eq. (1) accounts for the precession of the magnetization vector ($\vec{M}$) around $\vec{H}_{\mathrm{eff}}$. The second term with the first-order time derivative of $\vec{M}$, is the Gilbert damping term [24], which models the transfer of energy and angular momentum of $\vec{M}$ to the surrounding degrees of freedom (relaxation of $\vec{M}$ towards $\vec{H}_{\mathrm{eff}}$). Figure 4(a) shows the background subtracted time-resolved Kerr rotation data (precessional part) for two different values of the applied magnetic field, fitted with a damped sinusoidal function,

$$M(t) = M(0)e^{-t/\tau_d}\sin(2\pi f t) \quad (2)$$

Here, $M(0)$ is the initial amplitude of the precession, $\tau_d$ is the relaxation time of the coherent precession obtained as a fitting parameter, and $f$ is the precession frequency, which can also be extracted directly from the Fast Fourier Transform (FFT) of the precessional oscillation. Due to the size of the laser spot (D $\sim$ 500 $\mu$m), our measurement basically detects only magnon wavevectors up to approximately k $\sim \pi/500$ rad/$\mu$m, thus essentially only the ferromagnetic resonance (FMR). The effective magnetization ($M_{\mathrm{eff}}$), which includes the saturation magnetization and potential additional out-of-plane anisotropies, is calculated from the magnetic field dependence of the precession frequencies (Fig. 4(b)) and fitting the data points with the Kittel formula [31],

$$f = \frac{1}{2\pi}\sqrt{\omega_{\mathrm{H}}(\omega_{\mathrm{H}} + \omega_{\mathrm{M}})} \quad (3)$$

where $\omega_{\mathrm{H}} = \gamma\mu_0 H$, $\omega_{\mathrm{M}} = \gamma\mu_0 M_{\mathrm{eff}}$ and $H$ is the externally applied magnetic field and $\gamma = 1.83 \times 10^{11}$ rad s$^{-1}$ T$^{-1}$ for Py. Strictly speaking, Eq. 3 is only valid for a completely in-plane magnetized film, but we have verified using micromagnetic simulations that it approximates our experimental situation very well. From the fit, $M_{\text{eff}}$ is obtained to be $\sim 700 \pm 25$ kA m$^{-1}$ for 5 nm Py [and $\sim 670 \pm 20$ kA m$^{-1}$ for 2.8 nm Py, see Supplemental Materials] measured at $F = 4.6$ mJ cm$^{-2}$.

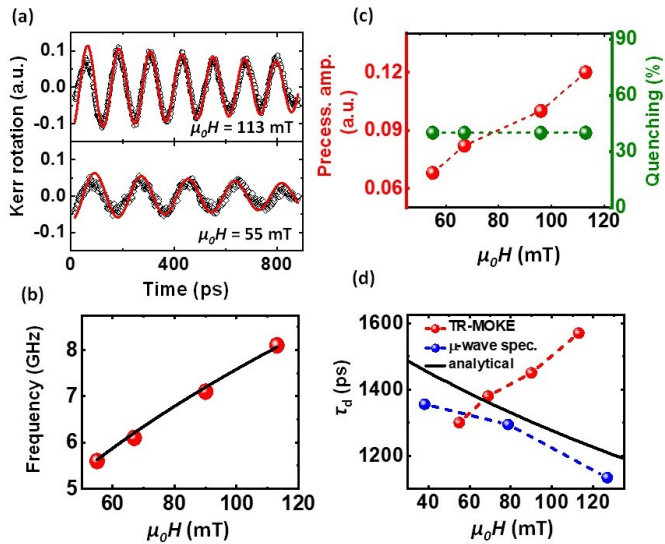

FIG. 4. Magnetic field dependent dynamics: (a) Background-subtracted time-resolved Kerr rotation data for 5 nm Py sample measured at two different magnetic fields and at $F = 4.6$ mJ cm$^{-2}$. Solid lines are fitting lines. (b) Magnetic field dependence of precession frequency. Solid line represent the Kittel fit to the data points. (c) Magnetic field dependence of quenching (green) and of precessional amplitude (red). (d) Magnetic field dependence of the precessional relaxation time ($\tau_d$) measured after ultrafast demagnetization in TR-MOKE (red), using microwave spectroscopy (blue) and analytical calculation (black).

When studying the excitation of the coherent magnons, points of interest are the dependence of the precession frequency, amplitude and lifetime on the external conditions. Concerning the amplitude, Fig. 4(c) shows the dependence of quenching and precessional amplitude on the magnetic field for a fixed pump fluence ($F = 4.6$ mJ cm$^{-2}$). As expected, the quenching, which is a measure of the energy initially introduced into the magnetic system, is independent of the applied magnetic field. Surprisingly at first glance, the precessional amplitude, which is usually considered as a measure for the energy of the coherent oscillations, is strongly dependent on magnetic field. We interpret the observed increase of the precession amplitude with increasing magnetic field as an increase of the part of the coherent precession that has a well-defined and constant phase relation to the pump pulse. This can be explained by the fact that the static out-of-plane component of the magnetization, which is required for TR-MOKE to measure the coherent precession [1, 2], increases with the strength of the applied bias magnetic field.

Interestingly, we also observe that the precessional relaxation time ($\tau_d$) as obtained from TR-MOKE measurements, increases with increasing magnetic field (red dots in Fig. 4(d)). This is unexpected if one assumes that Gilbert damping, described by a material parameter (Gilbert damping constant $\alpha$) is responsible for the decay of the precession. Since Gilbert damping is viscous, it predicts a decrease of the magnon lifetime with increasing frequency, which is equivalent to an increase of the magnetic field strength in the presented geometry. Analytical calculations of the magnon characteristics based on the Gilbert model [32–35] show that the Gilbert-induced lifetime decreases with increasing magnetic field (black solid line in Fig. 4(d)). For the analytical calculations, we have assumed an in-plane magnetic field, but we have verified again with micromagnetic simulations that this approximation is well justified. The lifetime ($\tau_d$) of the homogeneous FMR mode for an in-plane magnetized (the applied in-plane magnetic field is assumed to be $H$) thin film is calculated using the following expression [35]:

$$\frac{1}{\tau_d} = \alpha \left( \omega_{\text{H}} + \frac{\omega_{\text{M}}}{2} \right) \quad (4)$$

A possible interpretation for this intriguing discrepancy between the lifetimes measured with different excitation mechanisms could be the contribution of non-Gilbert damping mechanisms. One of these is the so-called two-magnon scattering mechanism, where the magnon energy is redistributed from the FMR ($k = 0$) to other short-wavelength magnons ($k > 0$) due to defect-induced scattering. Arias and Mills [17] developed a theory describing the contributions of two-magnon scattering to the FMR linewidth. The two-magnon process is linear in magnon amplitude, thus the broadening of the FMR linewidth is independent of the magnon amplitude. Woltersdorf et al., showed by TR-MOKE experiments that different capping layers can affect spin relaxation and damping of Fe films in different ways [19]. They also showed the increase of relaxation time or decrease of damping with increasing magnetic field for Cu capped Fe films and interpreted it by two-magnon scattering. Liu et al. [21] showed that the effective damping constant decreases with the increasing magnetic field, suggesting a contribution of magnetic anisotropy to the enhanced damping. Some other reports have also discussed the enhancement of damping with decreasing magnetic fields due to two-magnon scattering, magnetic anisotropy or spin pumping effects [11, 20, 22].

To clarify a possible contribution of two-magnon scattering from defects at surfaces and interfaces [17, 18] to the measured decay, we performed additional independent measurements of the FMR lifetime using inductive microwave spectroscopy measured with a vector network analyzer (VNA), as shown by the blue dots in Fig. 4(d). This technique is known to be sensitive to linewidth broadening induced by two-magnon scattering [36]. However, the measurements show a decrease of lifetime in accordance with the Gilbert model. In addition, a theoreti-cal work [18] predicts that a two-magnon contribution to the linewidth should increase with resonance frequency and magnetic field in our experimental case (small angles of less than 15° to the films plane). Thus, we can conclude that the defect induced two-magnon scattering is not responsible for the increase in lifetime ($\tau_d$) observed in our TR-MOKE measurements. Instead, we interpret the change in lifetime due to coherent precession as follows. Both incoherent and coherent magnons are excited when the sample is hit with a fs laser pulse in TR-MOKE. The ratio of coherent to incoherent magnons is influenced by the out-of-plane component of the static magnetization, which breaks the symmetry of the system. As in the case of the precession amplitude, a higher external field strength increases the static out-of-plane component and thus the relative proportion of coherent magnons with a defined and constant phase relationship to the laser pulses. Due to the reduced excitation of the incoherent magnons at higher fields, the dephasing of the precession signal is weaker, leading to a longer lifetime of the measured coherent precession signal. As the external magnetic field decreases, the relative proportion of incoherent magnons to coherent magnons increases, leading to a lower lifetime. This is in contrast to microwave spectroscopy measurements where only coherent magnons are excited and only a negligible amount of thermally excited incoherent magnons are present. This explains the different trends in the lifetime of magnons excited and measured by microwave spectroscopy and TR-MOKE.

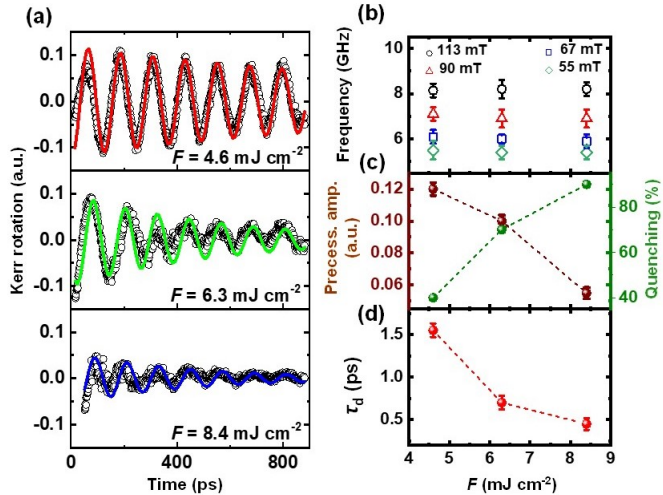

FIG. 5. Pump fluence dependent dynamics: (a) Background-subtracted time-resolved Kerr rotation data for 5 nm Py sample measured at $\mu_0 H = 113$ mT and different pump fluences. Solid lines are fitting lines. (b) Pump fluence dependence of precessional frequencies at $\mu_0 H = 113$ mT. (c) Pump fluence dependence of quenching (olive) and precessional amplitude (brown). Pump fluence dependent quenching plotted here (olive) is the same as plotted by solid circles in Fig. 2(e). (d) Pump fluence dependence of precessional relaxation time ($\tau_d$) measured after ultrafast demagnetization in TR-MOKE.

Another interesting parameter to study is the influence of the excitation intensity, which in our case is given by the pump fluence, on the coherent dynamics. The background subtracted time-resolved Kerr rotation data for the 5 nm Py sample measured at different pump fluences ($F$) are shown in Fig. 5(a). The energy deposited by the pump pulse, in the form of heat within the probed volume, plays a very crucial role in the modification of the local magnetic properties, i.e. the magnetic moment, anisotropy, coercivity, magnetic susceptibility, etc., as well as the precession frequency can experience a variation with the pump fluence[12, 37]. However, we do not observe any significant frequency shift within our experimental fluence range as shown in Fig. 5(b). This also indicates that the temperature of the sample is not significantly increased on longer timescales after the initial, fast remagnetization. Consequently, $M_{\text{eff}}$ calculated from Eqn. (3) shows no significant dependence on $F$ within our experimental fluence range. Thus, we conclude that as the pump fluence increases, there is no further change in the anisotropy field that can modify the effective magnetization of the system to this extent [21]. Figure 5(c) shows that the quenching increases with pump fluence as expected. However, although the initial quenching is a measure of the absorbed energy and thus the source of coherent precession, the precession amplitude decreases with fluence. The precessional relaxation time ($\tau_d$) also shows a decrease with pump fluence (Fig. 5(d)). The decrease of both the precession amplitude and the precession relaxation time can be explained by the dephasing of the magnons due to nonlinear magnon-magnon interactions. These interactions generally increase with an increase in the total magnon population (incoherent and coherent), which we believe is proportional to the quenching. In addition, the probability of excitation of incoherent magnons relative to coherent magnons increases with the increase of disorder in the system and thus with the quenching / pump fluence.

## IV. SUMMARY

In summary, the spin dynamics on different time scales in thin $Ni_{80}Fe_{20}$ films have been studied using an all-optical TR-MOKE technique. We study the precession dynamics in the GHz range after femtosecond laser-induced ultrafast demagnetization. The demagnetization time, fast remagnetization time, and magnetization quenching studied in both longitudinal and polar geometry show an increasing trend with excitation fluence, consistent with a spin-flip scattering-dominated demagnetization process. On a longer timescale of several hundreds of picoseconds, we observe an increase of the coherent precessional relaxation time with magnetic field / resonance frequency, which cannot be explained by viscous Gilbert damping. Using standard FMR techniques, we conclude that two-magnon scattering is not responsible for this behavior. Instead, we can consistently explain all observed trends by considering the different relative con-tributions of coherent and incoherent magnons produced in the ultrafast demagnetization process and the nonlinear interaction between them. This interpretation also explains the dependence of the coherent magnon amplitude and relaxation time on the excitation fluence. We expect that our results will pave the way for future experimental and theoretical investigations towards a deeper understanding of the photon-to-magnon conversion in ultrafast demagnetization processes.

## V. ACKNOWLEDGMENTS

The authors thank Eva Prinz, Jonas Hoefer, and Martin Stiehl for technical assistance. The work was funded by the Deutsche Forschungsgemeinschaft (DFG, German Research Foundation) under granz No. TRR 173-268565370 Spin+X: spin in its collective environment (project B11 and B03).

[1] G. Ju, A. V. Nurmikko, R. F. C. Farrow, R. F. Marks, M. J. Carey, and B. A. Gurney, Phys. Rev. Lett. **82**, 3705 (1999).
[2] M. van Kampen, C. Jozsa, J. T. Kohlhepp, P. LeClair, L. Lagae, W. J. M. de Jonge, and B. Koopmans, Phys. Rev. Lett. **88**, 227201 (2002).
[3] B. Koopmans, J. J. M. Ruigrok, F. D. Longa, and W. J. M. de Jonge, Phys. Rev. Lett. **95**, 267207 (2005).
[4] J. Walowski, G. Müller, M. Djordjevic, M. Münzenberg, M. Kläui, C. A. F. Vaz, and J. A. C. Bland, Phys. Rev. Lett. **101**, 237401 (2008).
[5] E. Beaurepaire, J.-C. Merle, A. Daunois, and J.-Y. Bigot, Phys. Rev. Lett. **76**, 4250 (1996).
[6] M. Vomir, L. H. F. Andrade, L. Guidoni, E. Beaurepaire, and J.-Y. Bigot, Phys. Rev. Lett. **94**, 237601 (2005).
[7] J. Bigot, M. Vomir, L. Andrade, and E. Beaurepaire, Chem. Phys. **318**, 137 (2005).
[8] M. Djordjevic and M. Münzenberg, Phys. Rev. B **75**, 012404 (2007).
[9] A. Barman, S. Wang, J. D. Maas, A. R. Hawkins, S. Kwon, A. Liddle, J. Bokor, and H. Schmidt, Nano Lett. **6**, 2939 (2006).
[10] J. Walowski, M. D. Kaufmann, , B.Lenk, C. Hamann, J. McCord, and M. Münzenberg, J. Phys. D: Appl. Phys. **41**, 164016 (2008).
[11] M. Djordjevic, G. Eilers, A. Parge, M. Münzenberg, and J. S. Moodera, J. Appl. Phys. **99**, 08F308 (2006).
[12] S. Mondal and A. Barman, Phys. Rev. Appl. **10**, 054037 (2018).
[13] S. Mukhopadhyay, S. Majumder, S. N. Panda, and A. Barman, Nanotechnol. **34**, 235702 (2023).
[14] J. Bigot and M. Vomir, Ann. Phys. (Berlin) **525**, 2 (2013).
[15] A. Barman and J. Sinha, Spin dynamics and damping in ferromagnetic thin films and nanostructures (Springer Cham, 2017) pp. 1–156.
[16] S. Parchenko, D. Pecchio, R. Mondal, P. M. Oppeneer, and A. Scherz, Arxiv **arXiv:2305.00259**, 1 (2023).
[17] R. Arias and D. L. Mills, Phys. Rev. B **60**, 7395 (1999).
[18] P. Landeros, R. E. Arias, and D. L. Mills, Phys. Rev. B **77**, 214405 (2008).
[19] G. Woltersdorf, M. Buess, B. Heinrich, and C. H. Back, Phys. Rev. Lett. **95**, 037401 (2005).
[20] G. Malinowski, K. C. Kuiper, R. Lavrijsen, H. J. M. Swagten, and B. Koopmans, Appl. Phys. Lett. **94**, 102501 (2009).
[21] B. Liu, X. Ruan, Z. Wu, H. Tu, J. Du, J. Wu, X. Lu, L. He, R. Zhang, and Y. Xu, Appl. Phys. Let. **109**, 042401 (2016).
[22] Y. Tserkovnyak, A. Brataas, and G. E. W. Bauer, Phys. Rev. Lett. **88**, 117601 (2002).
[23] R. Wilks, R. J. Hicken, M. Ali, B. J. Hickey, J. D. R. Buchanan, A. T. G. Pym, and B. K. Tanner, J. Appl. Phys. **95**, 7441 (2004).
[24] T. Gilbert, IEEE Trans. Magn. **40**, 3443 (2004).
[25] F. Dalla Longa, J. T. Kohlhepp, W. J. M. de Jonge, and B. Koopmans, Phys. Rev. B **75**, 224431 (2007).
[26] B. Koopmans, G. Malinowski, F. Dalla Longa, D. Steiauf, M. Fähnle, T. Roth, M. Cinchetti, and M. Aeschlimann, Nat. Mater. **9**, 259 (2010).
[27] T. Roth, A. J. Schellekens, S. Alebrand, O. Schmitt, D. Steil, B. Koopmans, M. Cinchetti, and M. Aeschlimann, Phys. Rev. X **2**, 021006 (2012).
[28] M. Krauß, T. Roth, S. Alebrand, D. Steil, M. Cinchetti, M. Aeschlimann, and H. C. Schneider, Phys. Rev. B **80**, 180407 (2009).
[29] M. Cinchetti, M. Sánchez Albaneda, D. Hoffmann, T. Roth, J.-P. Wüstenberg, M. Krauß, O. Andreyev, H. C. Schneider, M. Bauer, and M. Aeschlimann, Phys. Rev. Lett. **97**, 177201 (2006).
[30] S. Eich, M. Plötzing, M. Rollinger, S. Emmerich, R. Adam, C. Chen, H. C. Kapteyn, M. M. Murnane, L. Plucinski, D. Steil, B. Stadtmüller, M. Cinchetti, M. Aeschlimann, C. M. Schneider, and S. Mathias, Sci. Adv. **3**, e1602094 (2017).
[31] C. Kittel, Phys. Rev. **73**, 155 (1948).
[32] B. Kalinikos, IEE Proceedings H Microwaves, Optics and Antennas **127**, 4 (1980).
[33] B. A. Kalinikos and A. N. Slavin, J. Phys. C: Solid State Phys. **19**, 7013 (1986).
[34] B. A. Kalinikos, M. P. Kostylev, N. V. Kozhus, and A. N. Slavin, J. Phys.: Cond. Matt. **2**, 9861 (1990).
[35] D. D. Stancil and A. Prabhakar, Quantum theory of spin waves, in Spin Waves: Theory and Applications (Springer US, Boston, MA, 2009) pp. 33–66.
[36] I. Neudecker, G. Woltersdorf, B. Heinrich, T. Okuno, G. Gubbiotti, and C. Back, J. Magn. Magn. Mater. **307**, 148 (2006).
[37] S. Mizukami, H. Abe, D. Watanabe, M. Oogane, Y. Ando, and T. Miyazaki, Appl. Phys. Exp. **1**, 121301 (2008).

# Coherent and incoherent magnons induced by strong ultrafast demagnetization in thin permalloy films

Anulekha De,[1,*] Akira Lentfert,[1] Laura Scheuer,[1] Benjamin Stadtmüller,[1,2] Georg von Freymann,[1,3] Martin Aeschlimann,[1] and Philipp Pirro[1,†]

[1]*Department of Physics and Research Center OPTIMAS, Rheinland-Pfälzische Technische Universität Kaiserslautern-Landau, 67663 Kaiserslautern, Germany*
[2]*Institute of Physics, Johannes Gutenberg University Mainz, 55128 Mainz, Germany*
[3]*Fraunhofer Institute for Industrial Mathematics, 67663 Kaiserslautern, Germany*

## S1. Three Temperature Model:

The dynamics of the spin fluctuations after excitation by ultrafast laser pulses can be described by a phenomenological thermodynamic model, the so-called three-temperature model (3TM), which pictures how energy is redistributed among electrons, spins, and the lattice after the absorption of the laser power by the electronic system. The energy flow ultimately leads to an increase in the spin temperature, thereby reducing the magnetization. The expression is given by,

$$-\frac{\Delta M}{M} = \left\{ \left[ \frac{A_1}{(1+t/\tau_0)^{1/2}} - \frac{A_2\tau_E - A_1\tau_M}{\tau_E - \tau_M} e^{-t/\tau_M} - \frac{\tau_E(A_1 - A_2)}{\tau_E - \tau_M} e^{-t/\tau_E} \right] H(t) + A_3\delta(t) \right\} \otimes G(t)$$

Here, $\tau_M$ and $\tau_E$ are ultrafast demagnetization and first remagnetization times respectively. $A_1$ represents the amplitude of magnetization after fast relaxation, $A_2$ is proportional to the maximum electron temperature rise, and $A_3$ represents the state filling effects during pump-probe temporal overlap. $H_S(t)$ is the Heaviside step function, $\delta(t)$ is the Dirac delta distribution, and $G(t)$ is a Gaussian function corresponding to the laser pulse. $\tau_0$ represents the cooling time through heat diffusion. This model is very useful in analysing experimental data and extracting quantitative information on the timescales of the different processes taking place during the laser induced ultrafast demagnetization.

## S2. Precession dynamics for 2.8 nm Py sample:

Figure S1(a) shows the background subtracted time-resolved Kerr rotation data (precessional part) measured at different values of magnetic fields, fitted with a damped sinusoidal function (Eqn. 2 of the main article) for 2.8 nm Py sample. All measurements are done at particular pump fluence of F = 4.6 mJ cm$^{-2}$. Figure S1(b) shows magnetic field dependence of the precession frequencies obtained from the fast Fourier transform (FFT) of the precessional oscillation, from which we calculate the effective magnetization ($M_{eff}$) using the Kittel formula (Eqn. (3) of the main article). The value of $M_{eff}$ obtained from fit is 670 ± 20 kA m$^{-1}$, which is slightly less than the thicker (5 nm Py) film. Figure S1(c) shows the variation of precessional relaxation time ($\tau_d$) with magnetic as obtained after ultrafast demagnetization in TR-MOKE measurements (red), microwave spectroscopy measurements (blue) and analytical calculations (black). We observe that $\tau_d$ increases with magnetic field for TR-MOKE measurements which is in contrast to both the microwave spectroscopy measurements and analytical calculations. Similar results are obtained from 5 nm Py sample and are thoroughly discussed in the main article.

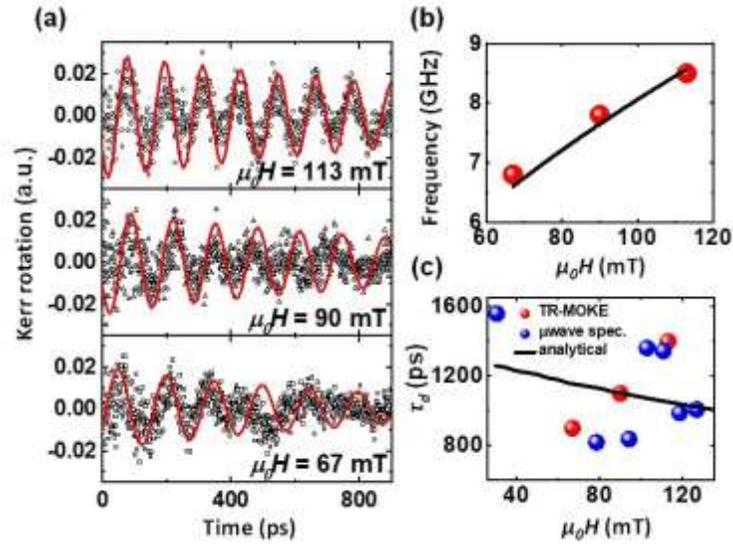

**Fig. S1:** Magnetic field dependent dynamics of 2.8 nm Py sample: (a) Background-subtracted time-resolved Kerr rotation data measured at different values of magnetic field and at F = 4.6 mJ cm$^{-2}$. Solid lines are fitting lines. (b) Magnetic field dependence of precession frequency. Solid line represent the Kittel fit to the data points. (c) Magnetic field dependence of precessional relaxation time ($\tau_d$) measured after ultrafast demagnetization in TR-MOKE (red), microwave spectroscopy (blue) and analytical calculation (black).

### S3. Effect of in-plane magnetic field:

For the in-plane configuration of the magnetic field, we observe a reduction in precessional amplitude leading to a poor signal-to-noise ratio (as shown in S2(a)). As the magnetic field is tilted slightly (~ 15°) in the out-of-plane direction, the precessional amplitude increases and we observe a clear time-resolved Kerr rotation trace resulting clear fast Fourier transformed magnon modes. In the in-plane configuration the dominance of nonmagnetic noise due to two-magnon scattering has suppressed the features of magnetic peaks and only a few spurious peaks are present in the spectra. A comparison between the FFT powers for in-plane and tilted magnetic fields are shown in Fig. S2(c), where the power is negligibly small in in-plane configuration as compared to the tilted configuration of magnetic field.

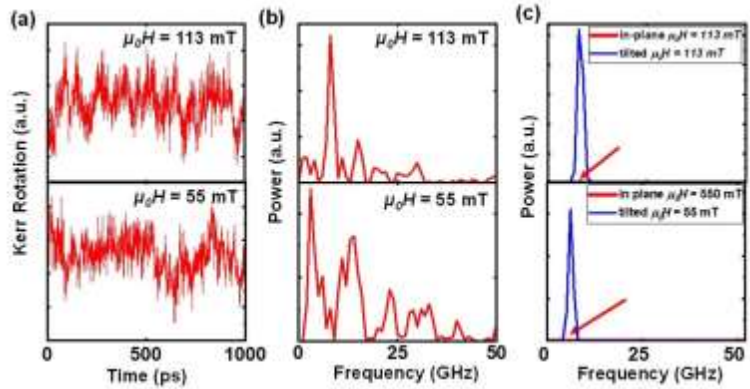

**Fig. S2:** (a) Precessional dynamics at two different values of in-plane magnetic field. (b) The corresponding FFT power spectra (c) The comparison between FFT spectra at in-plane and tilted magnetic fields. The values of magnetic fields are mentioned in the respective graphs.